\documentclass{emulateapj}

\slugcomment{Recieved 2006 May 17; accepted 2006 August 7; to be published in ApJ letters}

\shorttitle{The water vapor abundance in Orion~KL outflows}
\shortauthors{Cernicharo et al.}

\begin{document}

\title{The water vapor abundance in Orion KL outflows\thanks{Based on observations 
with ISO, an ESA project with instruments funded by ESA Member States
(especially the PI countries: France, Germany, the Netherlands
and the United Kingdom) and with participation of ISAS and NASA.}}

\author{Jos\'e Cernicharo\altaffilmark{2}, 
Javier R. Goicoechea\altaffilmark{3}, Fabien Daniel\altaffilmark{2},\\ 
Mercedes R. Lerate\altaffilmark{4,5}, 
Michael J. Barlow\altaffilmark{5}, 
Bruce M. Swinyard\altaffilmark{4}, \\
Ewine van Dishoeck\altaffilmark{6}, 
Tanya L. Lim\altaffilmark{4}, 
Serena  Viti\altaffilmark{5}, 
Jeremy Yates\altaffilmark{5}}

\altaffiltext{2}{DAMIR, Instituto de Estructura de la Materia, Consejo
Superior de Investigaciones Cient\'{i}ficas, 
Serrano 121, 28006, Madrid, Spain. (cerni@damir.iem.csic.es) }
\altaffiltext{3}{LERMA, UMR 8112, CNRS, Observatoire de Paris and 
Ecole Normale Sup\'erieure, 24 Rue Lhomond, 75231 Paris 05, France.}
\altaffiltext{4}{Rutherford Appleton Laboratory, Chilton, 
Didcot,  UK. }
\altaffiltext{5}{University College London, Gower Street, 
London WC13, UK.} 
\altaffiltext{6}{Sterrewacht Leiden, PO Box 9513, 2300 RA Leiden, The Netherlands.}

\begin{abstract}

We present the detection and modeling of more than 70 far--IR pure rotational
lines of water vapor, including the $^{18}$O and $^{17}$O isotopologues,
towards Orion~KL. Observations were performed with the \textit{Long
Wavelength Spectrometer} Fabry--Perot (LWS/FP; $\lambda/\Delta \lambda
\sim$6800-9700) on board the \textit{Infrared Space Observatory} (ISO)
between $\sim$43 and 197~$\mu$m. The water line profiles evolve
from P--Cygni type profiles (even for the H$_{2}^{18}$O lines) to pure
emission at wavelengths above $\sim$100~$\mu$m. We find that most of the 
water emission/absorption arises from an extended flow of gas expanding at
25$\pm$5~km~s$^{-1}$. Non--local radiative transfer models show that much
of the water excitation and line profile formation is driven by the dust
continuum emission. The derived beam averaged water abundance
is 2--3$\times$10$^{-5}$. The inferred gas temperature $T_k$=80--100~K
suggests that: $(i)$ water could have been formed in the \textit{plateau}
by gas phase neutral-neutral reactions with activation barriers if the gas was previously
heated (e.g.~by shocks) to $\geq$500~K and/or $(ii)$ H$_2$O formation in the
outflow is dominated by \textit{in--situ} evaporation of grain water--ice
mantles and/or $(iii)$ H$_2$O was formed in the innermost and warmer
regions (e.g. the hot core) and was swept up in $\approx$1000~yr, the
dynamical timescale of the outflow. 

\end{abstract}


\section{Introduction}

Star forming regions are associated with violent phenomena
such as cloud collapse, molecular outflows and related shocked regions.
Under these conditions the neutral gas is warm and  water vapor is predicted to be
abundant \citep{dra82} and to play a dominant role in the thermal balance
\citep{neu93}. 
Unfortunately, ground--based observations of H$_2$O
are difficult and so the determination of water column densities 
is not straightforward. 
Nevertheless, ground--based observations of water maser lines have been performed
toward Orion~KL. From VLBI observations of the 6$_{16}$--5$_{23}$ line
at $\sim$22~GHz, \citet{gen81} determined the kinematics and the
expansion velocity for the so--called
\textit{low velocity outflow} ($\sim$18$\pm$2~km~s$^{-1}$). 
The widespread nature of 
water vapor has been probed with maps of the
3$_{13}$--2$_{20}$ line at $\sim$183~GHz \citep{cer90,cer94}.
This was the first time that the water abundance
was estimated in the different large scale components around Orion~KL. 
In addition, the high excitation conditions of the \textit{plateau} gas (a mixture
of  outflows, shocks and interactions with the ambient cloud) has been 
revealed by observations of the 5$_{15}$--4$_{22}$ line
at $\sim$325~GHz \citep{men90,cer99b}. 
Due to their maser nature, the water abundance determination 
from these lines is quite involved. Even the observation of the 
H$_{2}^{18}$O 3$_{13}$--2$_{20}$ line
at $\sim$203~GHz gives only poor estimates due to the
overlap with other molecular lines \citep{jac88}.
The HDO species has been also detected and modeled towards Orion~KL
\citep{jac90,par01}, but the specific gas and dust chemistry
has to be taken into account to derive $\chi$(H$_2$O).

ISO \citep{kes96} spectrometers have provided the opportunity to observe
many IR H$_2$O lines towards bright sources such as 
Orion~KL. In particular, Harwitt et al. (1998) presented eight ISO/LWS/FP
\citep{cle96} 
water lines between $\sim$71 and $\sim$125~$\mu$m involving 
energy levels between 300 and 800~K. They
estimated $\chi$(H$_2$O)$\simeq$5$\times$10$^{-4}$;
however, the 
interpretation of these lines should include
radiative pumping from IR dust photons.
A larger set of weaker IR lines,
including
those of H$_{2}^{18}$O and H$_{2}^{17}$O, is needed to minimize
the opacity effects always associated with H$_{2}^{16}$O lines.
Orion~KL has also been targeted with the
\textit{Short Wavelength Spectrometer} (SWS) tracing a smaller
region than that
probed by the LWS beam below $\sim$45~$\mu$m \citep{vd98}. 
Nineteen water absorption lines with energies
between 200 and 750~K were detected \citep{wri00}. These authors 
suggested that mid--IR water lines
arise from the \textit{low--velocity outflow} and estimated  
$\chi$(H$_2$O)=2--5$\times$10$^{-4}$ (for an assumed gas temperature
of 200--350~K). 
Most recently, the  1$_{10}$--1$_{01}$ 
lines of H$_{2}^{16}$O and H$_{2}^{18}$O 
have been observed with SWAS 
($\chi$(H$_{2}^{16}$O)$\simeq$3.5$\times$10$^{-4}$; \cite{mel00}; \cite{sne00})
and ODIN ($\chi$(H$_{2}^{16}$O)$\simeq$10$^{-5}$--10$^{-4}$; \cite{olo03}). 
In these observations, the H$_{2}^{16}$O ground--state line shows
widespread emission over a $\sim$5$'$ region.

In this letter we present all water lines detected
in the first far--IR line survey towards Orion~KL carried out with
the ISO/LWS spectrometer
(Lerate et al. 2006), and the radiative transfer
models that fit the water data set (involving levels up to 2000 K).

\begin{figure*} [ht] 
\centering
\includegraphics[angle=0,width=14cm]{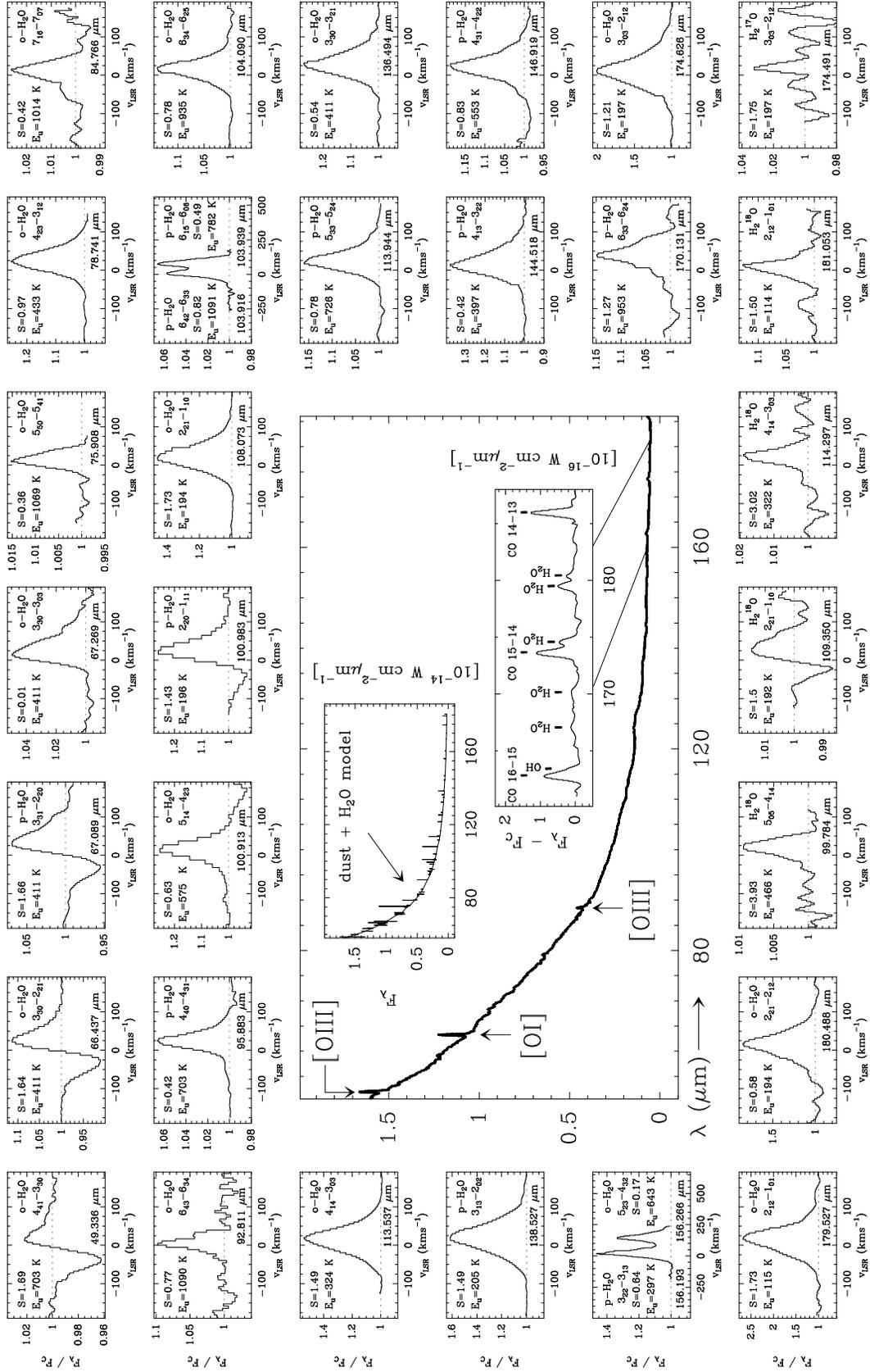}
\caption{Summary of the 70 far--IR water lines detected by the ISO/LWS-FP 
towards Orion KL. 
The ordinate corresponds to the continuum normalized flux and the 
abscissa to the velocity (in km~s$^{-1}$). Transition rotational 
numbers, rest wavelengths (in $\mu$m), upper level energies (in K) and 
intrinsic line strengths are shown in each box. The central inset shows 
the ISO/LWS-grating observations towards Orion at a resolution of 
$\lambda/\Delta \lambda \sim 300$. The ordinate corresponds to the
absolute flux (in W~cm$^{-2}$~$\mu$m$^{-1}$) and the 
abscissa to the wavelength (in $\mu$m).
Main molecular features  between $\sim$160 and  
$\sim$197~$\mu$m are labelled in a zoom to the grating spectrum. 
The full radiative transfer model for the far--IR continuum and
water line spectrum is also shown in the upper inset.}
\label{fig-observations}
\end{figure*}

\clearpage

\section{Observations and Data reduction}
All water lines within the range of the ISO/LWS
($\sim$43-197$\mu$m) were
observed using long integrations with the FP spectrometer (L04 AOT) which
provides the largest spectral resolution for the LWS instrument.
Preliminary results were presented by \citet{cer99a}.
In addition, a complete less sensitive (L03 AOT)
far--IR line survey of Orion has been
carried out.
Adding both data sets more than 70 water lines have been detected 
(Lerate et al. 2006). 
The LWS circular aperture size is $\sim$80$''$, although it slightly
depends on the particular LWS detector.
In its FP mode  the instrumental response is close to a broad--wing Lorentzian
with a spectral resolution of $\lambda/\Delta \lambda \sim$6800-9700.
Processing of the water lines from  AOT L03 was carried out using the 
Offline Processing (OLP) pipeline and the LWS Interactive Analysis (LIA)
package version 10. 
AOT L04 spectra were analyzed using the ISO Spectral Analysis Package 
(ISAP\footnote{ISAP is a joint development by the LWS and SWS Instruments
Teams and Data Centers. Constributing institutes are CESR, IAS,
IPAC, MPE, RAL and SRON.}).
The full description
of the complex data calibration and reduction process and  associated 
target dedicated time numbers, and all the tabulated line observations
and spectroscopic parameters are given by Lerate et al. (2006).

\section{Results and discussion}
A summary of the resulting water lines is shown in Fig.~\ref{fig-observations}.
From these far--IR observations it is clear that water lines
show a complicated behavior.
For wavelengths above $\sim$100~$\mu$m,
H$_{2}^{16}$O  lines  are observed in emission. 
However, for shorter wavelengths, lines arising from energy levels below 
$\sim$1000~K and with large line--strengths, show P--Cygni profiles with 
emission covering a large velocity
range. However, those with weak line-strengths or arising from higher energy
levels, are observed in pure emission.
H$_{2}^{18}$O lines also show P--Cygni type profiles 
below $\sim$100~$\mu$m. In the H$_{2}^{18}$O  case, the absorption 
component is deeper than in the analogous H$_{2}^{16}$O transition. 
Since optical depth effects  are much less severe in H$_{2}^{18}$O lines, 
the associated
P--Cygni type profiles trace the main origin of far--IR water lines
toward Orion~KL,
i.e. an extended outflow.
Pure emission H$_2$O lines peak 
around $v\simeq$20~km~s$^{-1}$ (the source $v_{LSR}$ is 
$\sim$9~km~s$^{-1}$; Scoville et al. 1993) but it is likely 
that the most opaque lines are redshifted due to self--absorption. 
On the other hand, water lines detected below $\sim$50~$\mu$m 
are observed in pure absorption with a velocity peak of 
$v\simeq-$10~km~s$^{-1}$ (the same applies to most water
lines observed by ISO/SWS; \citet{wri00}). 

The turnover point between absorption and emission  lines is an
important clue to interpret this large data set and to determine
the relations between continuum+line opacity, line strengths, spatial 
distribution of gas, and physical conditions.
ISO observations clearly show
that most of the water vapor detected in the IR arises from a flow
of gas expanding at 25$\pm$5~km~s$^{-1}$. 
The inferred expansion velocity
is consistent with the \textit{low--velocity outflow} originally
revealed by $\sim$22~GHz  H$_2$O maser motions \citep{gen81}, but a contribution
from the extended \textit{high--velocity outflow} could be present
\citep{cer94}. However, no high velocity line--wing emission
is observed at ISO's sensitivity and S/N.
Similar conclusions have been found for the OH 
excited rotational lines \citep{goi06}.

The main problem with modeling ISO data is the limited spatial resolution,
which makes difficult to constrain the size and origin of the water region.
Besides, any detailed fit to the data requires
a detailed knowledge of the geometry and of the relative distribution of
dust continuum and water lines. Fortunately,  high angular resolution maps of water
at $\sim$183~ and $\sim$325~GHz lines do reveal the spatial and velocity
distribution of water in Orion \citep{cer94,cer99b}.
In particular, Cernicharo and coworkers detected at least four different
water components:
the extended  \textit{ridge} (extended quiescent gas), the \textit{plateau}
(including the \textit{high--} and \textit{low--velocity
outflows}) and the very narrow and strong features associated
with the small water bullets observed at $\sim$22~GHz \citep{gen81}.
Besides, the newly detected far--IR water lines associated with the highest 
energy levels may have a contribution from the \textit{hot core} 
(dense and hot inner regions). However,
the large far--IR line--plus--continuum opacity  will probably hide most of
the
\textit{hot core} emission. 
Finally,  radiative transfer effects in the most opaque lines,
e.g. self--absorption and/or scattering by a  lower density diffuse halo 
may possibly occur at velocity scales not resolved by ISO.\\

In order to estimate the water abundance, and the physical conditions
prevailing in the expanding gas,
we have modeled the first 30 rotational levels of both
ortho-- and para--H$_2$O  using the same
nonlocal code for lines and dust continuum as in the analysis of H$_2$O 
toward Sgr~B2 \citep{cer06}. The code has been described elsewhere
\citep{gon93} and has been recently 
improved to take into  account a more sophisticated 
description of the dust emissivity and radiative transfer 
(Daniel et al. 2006, in prep.).
The dust continuum emission has a crucial role in the radiative excitation
of light species such as H$_2$O or OH with transitions in the far--IR 
and has to be correctly taken into account.
In our model, level populations are consistently computed in statistical 
equilibrium  considering collisional excitation and pumping by line and
continuum photons. Collisional rates were scaled from those of H$_2$O--He 
collisions \citep{gre93}. A three--component model resembling
the \textit{hot core}, the  \textit{plateau} and the \textit{ridge} is adopted. 
In this work we assume that most of the  far--IR water lines arise only
from the \textit{plateau}. Of course, a minor contribution in the
lowest excitation water lines may come from the more extended regions
(producing narrow line emission) that we don't model here.
A low water abundance of (1--8)$\times$10$^{-8}$ has been estimated in these
regions from SWAS and ODIN observations (\cite{sne00}; \cite{olo03}) and thus,
the expected contribution to far--IR H$_2$O lines is small.
In order to have the closest view of the IR radiation field 
seen by water vapor we have simultaneously tried to reproduce the full 
SWS/LWS/ISO  continuum emission between $\sim$10 and $\sim$197~$\mu$m.

We find that the continuum level for 
$\lambda >$80~$\mu$m is well reproduced  by considering an inner 10$''$ 
IR continuum source (the \textit{hot core})  simulated by a grey--body, 
with a color temperature of 200--250K, which is optically thick in the 
far--IR ($\tau_{\lambda}=10\times 150/\lambda$).
The dust continuum emission from this  \textit{hot core} is 
attenuated by the surrounding components that we now describe.
Taking into account the spatial extent of the 183 and 325~GHz
water emission observed at higher spatial resolution \citep{cer90,cer94,cer99b},
 we model the \textit{plateau}
as a $\sim$40$''$ diameter unresolved shell expanding at 25$\pm$5~km~s$^{-1}$, where 
water molecules and dust grains coexist (T$_d$=T$_k$ is assumed).
The radiative effects caused by an additional, only dust,
5$'$ diameter component (the \textit{ridge}), with $T_d\simeq$25~K and 
$n$(H$_2$)$\simeq$10$^5$~cm$^{-3}$,  have been also included to fully match
the SWS/LWS continuum level.

\begin{figure} [t] 
\centering
\includegraphics[angle=0,width=8.5cm]{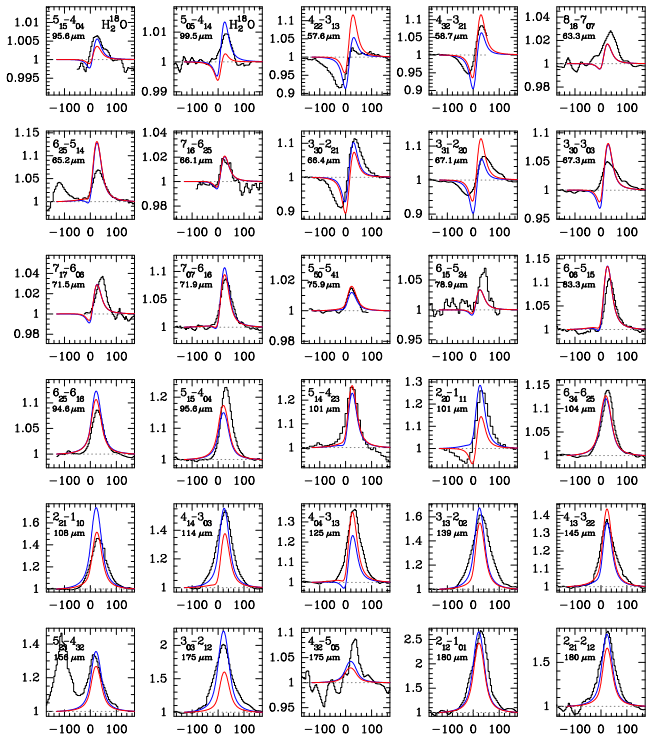}
\caption{Radiative transfer model results  discussed in the text.
Selected synthetic line profiles (continuous curves) are shown
over the observed ISO/LWS/FP detections (histograms). 
Model parameters are $\chi$(H$_2^{16}$O)=3$\times$10$^{-5}$,
$n(H_2)$=3.5$\times$10$^{5}$~cm$^{-3}$ and T$_k$=80~K (blue)
and $\chi$(H$_2^{16}$O)=2$\times$10$^{-5}$,
$n(H_2)$=2.5$\times$10$^{5}$~cm$^{-3}$ and T$_k$=100~K (red).
The ordinate corresponds to the continuum normalized flux and the 
abscissa to the velocity (in~km~s$^{-1}$).
Synthetic line profiles have been convolved with
a gaussian beam of 80$''$  and with a spectral resolution
characterized by a Lorentzian with a width of 33~km~s$^{-1}$.}
\label{fig-models}
\end{figure}

 Dust absorption coefficients 
are computed from tabulated optical grain properties from  Draine \& Lee (1984),
both in the \textit{plateau} and in the \textit{ridge}.
Due to the large optical depth of H$_2^{16}$O lines in the \textit{plateau}, 
we have first modeled the H$_2^{18}$O lines to obtain a more accurate
estimate  of the water abundance with less opacity constraint and 
faster level population convergence. 
A $^{16}$O/$^{18}$O abundance ratio of 500 has been adopted. 
After a reasonable fit to the H$_{2}^{18}$O lines, we iteratively
search for the best physical conditions that enable to 
simultaneously  reproduce the H$_{2}^{16}$O and H$_{2}^{18}$O
spectra. Fig. \ref{fig-models} shows two representative models
that qualitatively and quantitatively reproduce the majority of observed
line  profiles and intensities. The best models are found for
a \textit{plateau} temperature, density and water vapor abundance of 
$T_k$=80--100~K, $n$(H$_2$)=2.5--3.5$\times$10$^5$~cm$^{-3}$ 
and  $\chi$(H$_2^{16}$O)=2--3$\times$10$^{-5}$ respectively.
The physical conditions inferred from water lines in the \textit{plateau}
are similar to those obtained from OH lines \citep{goi06}, which suggests that
far--IR lines from both species trace the same expanding gas.
The derived water vapor abundance,
$\gtrsim$2$\times$10$^{-5}$, is obviously an averaged value
over the large LWS/ISO beam and  probably indicates that water can 
be locally more abundant, e.g. in the warmer interaction surfaces where the 
expanding gas  shocks the dense ambient material.
Larger angular resolution is needed to resolve
the  relative continuum and H$_2$O spatial  distribution
over this complex region. Nevertheless, the inferred 
gas temperatures, $\sim$100~K, are 
significantly below the gas temperature ($\sim$300~K) 
required to activate  the gas--phase  neutral--neutral  reactions
converting most of the available oxygen into water 
(abundances larger than $\sim$10$^{-4}$ are then predicted).
Taken into account the short dynamical timescale of the outflow, 
$\sim$1000~years, water vapor could have been formed by these 
neutral--neutral reactions if the gas in the \textit{plateau}
was previously (or is locally) heated to $\geq$500~K, e.g. by a C--shock passage 
\citep{ber98}.
However, if the \textit{plateau} gas temperature has reached a maximum temperature
of only $\sim$100~K, other formation mechanisms
are required to explain the observed water vapor abundances. 
In particular, the temperature inferred from ISO observations
is similar to the water--ice evaporation temperature, and
therefore,  \textit{in--situ} evaporation of  water--ice grain mantles formed in an 
evolutionary stage  prior to the onset of the outflow(s) 
could now dominate the  water vapor formation in the \textit{plateau}.
Finally, the observed H$_2$O could have been also  formed in the
innermost and warmest regions (e.g. the hot core) and had been swept up by the outflow.

To conclude, despite the large number of detected far--IR excited water
lines, neither  high gas temperature nor high density conditions
are required to populate the higher--energy water rotational levels.
Radiative pumping due to the strong IR radiation field
from the cloud is enough to populate these levels, at least
in the average picture given by ISO observations.
Future observations with  \textit{Herschel}
will allow one to map  many far--IR water resolved lines with improved 
angular resolution. 
It will then be possible to study in great detail
the role of water in star forming regions.

\acknowledgments
We are grateful to the LWS teams for the quality of the provided
data. We thank our refereree for useful comments and
suggestions.
We thank the Spanish DGES and PNIE grants AYA2000-1784,
ESP2001-4516 and AYA2003-2785.
JRG is supported by a \textit{Marie Curie Intra-European Individual
Fellowship} within the MEIF-CT-2005-515340 European Community FP6 contract.

\clearpage


\end{document}